\documentclass[amssymb,prl,aps,twocolumn,amsmath,showpacs]{revtex4}

\usepackage[dvips]{graphicx}
\usepackage{amstext}

\begin{document}

\title{Amplitude control of spin-triplet supercurrent in S/F/S Josephson junctions}
\author{William Martinez, W.P. Pratt, Jr., and Norman O. Birge}
\email{birge@pa.msu.edu} \affiliation{Department of Physics and
Astronomy, Michigan State University, East Lansing, Michigan
48824, USA}

\date{\today}

\begin{abstract}

Josephson junctions made with conventional s-wave superconductors and containing multiple layers of ferromagnetic materials can carry spin-triplet supercurrent in the presence of certain types of magnetic inhomogeneity.  In junctions containing three ferromagnetic layers, the triplet supercurrent is predicted to be maximal when the magnetizations of adjacent layers are orthogonal, and zero when the magnetizations of any two adjacent layers are parallel.  Here we demonstrate on-off control of the spin-triplet supercurrent in such junctions, achieved by rotating the magnetization direction of one of the three layers by 90$^{\circ}$.  We obtain ``on-off" ratios of 5, 7, and 19 for the supercurrent in the three samples studied so far.  These observations directly confirm one of the most salient predictions of the theory, and pave the way for applications of spin-triplet Josephson junctions in the nascent area of ``superconducting spintronics."

\end{abstract}

\pacs{74.50.+r, 74.45.+c, 75.70.Cn} \maketitle

When a superconducting (S) material is placed in contact with a non-superconducting material, the properties of both materials are modified close to the interface.  This ``superconducting proximity effect" can extend over distances of several hundred nanometers into the non-superconducting material at low temperatures \cite{Deutscher:69}.  When the non-superconducting material is ferromagnetic (F), in contrast, the proximity effect decays over a very short distance, of order one nm in strong F materials such as Fe or Co \cite{BuzdinReview:05,Robinson:06}.  This is because the electrons in a conventional superconductor have spin-singlet pairing symmetry; when such a pair enters a ferromagnetic material, one electron enters the majority spin band and the other enters the minority band.  Those two bands have different Fermi momenta, hence the pair acquires a finite momentum, or equivalently, the pair correlation function oscillates rapidly in space \cite{Demler:97}.  Those oscillations dephase equally rapidly in diffusive systems, leading to a very short decay length of the pair correlations in F.

In 2001, Bergeret et al. showed that a new type of proximity effect can arise in S/F systems in the presence of suitable forms of magnetic inhomogeneity near the S/F interface \cite{Bergeret:01}.  Specifically, the presence of non-collinear magnetizations can induce conversion of spin-singlet Cooper pairs from the conventional superconductor into spin-triplet pairs with projection $\pm$1 along the magnetization axis.  Such pairs consist of two electrons in the same spin band, hence the pairs do not feel the exchange energy, and they can penetrate deep into the ferromagnetic material \cite{Bergeret:01,Volkov:03,Eschrig:03,Bergeret:05}.  Experimental evidence for such spin-triplet pairs was first found in 2006 \cite{Keizer:06,Sosnin:06}; stronger evidence was then found by several groups in 2010 \cite{Khaire:10,Robinson:10,Sprungmann:10,Anwar:10,Wang:10}, using different approaches to produce the required non-collinear magnetization.

While the existence of spin-triplet pair correlations in ferromagnetic materials is no longer in doubt, the ability to control their amplitude has been slow to develop.  Our original work \cite{Khaire:10} was based on ``sandwich-style" Josephson junctions with the structure S/F'/N/SAF/N/F''/S, where F' and F'' are thin ferromagnetic layers, N are non-magnetic spacers, and SAF is a ``synthetic antiferromagnet" - in our case a Co/Ru/Co trilayer - which minimizes magnetic flux in the junction \cite{Khasawneh:09}.  Such a structure optimizes production of spin-triplet pairs when the magnetizations of adjacent layers in the structure are orthogonal to each other \cite{Houzet:07}.  In our original work the magnetic layers were multi-domain, so that the required noncollinear magnetization occurred randomly.  Later we found that magnetizing the samples resulted in a large increase in the critical supercurrent \cite{Klose:12}, because the SAF undergoes a spin-flop transition whereby the Co layers end up pointing in directions orthogonal to the direction of the F' and F'' layer magnetizations.  Although that enhancement was impressive, it cannot truly be called ``control" since the process could be reversed only by warming up the sample to room temperature, where the thin F' and F'' layers demagnetize.  Further progress in controlling the amplitude of spin-triplet supercurrent was made by Banerjee \textit{et al.} \cite{Banerjee:14} using S/F'/N/F/N/F''/S junctions and by Iovan \textit{et al.} \cite{Iovan:14} using asymmetric S/F'/N/F/S junctions.  Those groups found evidence for spin-triplet generation occurring during the magnetization reversal process while sweeping an external magnetic field.  Better control of the magnetic states has been achieved recently by several groups measuring the critical temperature $T_c$ of S/F/F  trilayers, where generation of spin-triplet correlations results in lowering of $T_c$ \cite{Leksin:12,Zdravkov:13,Wang:14,Jara:14,Flokstra:15,Singh:15}.  Controlling $T_c$, however, is less likely to be useful for future device applications than controlling supercurrent.

The goal of this work is to design a Josephson junction where the spin-triplet supercurrent can be controllably turned on and off by an external magnetic field and where these configurations persist when the field is removed.  To achieve this goal, we utilized a combination of ``hard" (F') and ``soft" (F'') ferromagnetic materials with vastly different switching fields, with the Co/Ru/Co SAF in between them.  From our previous work \cite{Klose:12,Wang:12}, we know that thin layers of Ni inside our Josephson junctions are quite hard, i.e. they have a large coercive field, $\mu_0 H_c \approx$ 50 mT. Ni is also very effective at producing spin-triplet supercurrent in S/F'/SAF/F''/S junctions \cite{Klose:12,Wang:12}.  For the soft layer, we chose the Permalloy alloy Ni$_{0.81}$Fe$_{0.19}$, which typically has $\mu_0 H_c$  of only a few mT.  The samples used in this work have F' = Ni(1.2nm) and F'' = NiFe(1.0nm).

Our first task was to determine the hardness of our Co/Ru/Co SAF - in other words, at what value of the external magnetic field does the SAF undergo the spin-flop transition whereby the remanent magnetization directions of the Co layers rotate by 90$^{\circ}$.  Standard magnetization measurements were not sufficient for this task, because the $M$ vs $H$ curves of our SAFs do not show any feature at the spin-flop transition; they are essentially linear until the SAF magnetization saturates at high field.  Instead we used the anisotropic magnetoresistance (AMR) to determine how the Co layers in the SAF respond to an external magnetic field \cite{Martinez:15}.  As expected, AMR measurements showed that samples with thinner Co layers are less sensitive to the applied field, but very thin Co won't suppress the short-range spin-singlet supercurrent adequately relative to the long-range spin-triplet supercurrent.  As a compromise, we chose to work with $d_\textrm{Co}$ = 4 nm (for a total Co thickness of 8 nm) in our actual Josephson junctions.  Josephson junctions containing such SAFs exhibit significantly suppressed spin-singlet supercurrent \cite{Khasawneh:09}, while SAF samples with $d_\textrm{Co}$ = 4 nm showed minimal change in AMR for applied fields less than 20 mT \cite{Martinez:15} -- the field range relevant to the experiments to be presented here.

A second important consideration in the sample design was the lateral size of the Josephson junctions. It is well known that the critical current, $I_c$, of a Josephson junction exhibits a ``Fraunhofer pattern" as a function of magnetic field applied in a direction perpendicular to the current flow.  For circular junctions with the current flowing out-of-plane and the field applied in-plane, $I_c$ follows an Airy pattern in flux   with its first minimum at $\Phi / \Phi_0$ = 1.22, where $\Phi_0$ = h/2e is the flux quantum.  The effective magnetic flux through the junction area,   $\Phi=\mu_0 Hw(2\lambda_L+d_N+d_F)+ \mu_0 Mwd_F$, includes the contribution from the external field $H$ and from the internal magnetization $M$, assuming the latter is uniform and collinear with $H$.  Here $w$ is the sample diameter, $d_N$ and $d_F$ are the thicknesses of the N and F layers in the junction, and the London penetration depth, $\lambda_L$, appears in the first term due to penetration of the external field into the top and bottom S electrodes.  The width of the central lobe of the Airy pattern in magnetic field is inversely proportional to the sample diameter $w$, and the central peak is displaced in the direction opposite to the direction of $M$ \cite{Ryazanov:99,Khaire:09}.  Since our experiment will involve changing the magnetization direction of the soft NiFe layer in our samples, we anticipate shifts in the Airy pattern central peak position during the course of the experiment.  While such shifts may be useful in their own right \cite{Held:06,Larkin:12}, from the point of view of this project they are a nuisance.  To avoid this complication, we must make our Josephson junctions sufficiently small so that the widths of the Airy patterns are much larger than any displacements of the central peak position.  For this experiment we fabricated samples with diameters of 0.5, 0.7, and 1.0  m.  Given that $\lambda_L$= 85 nm for our Nb, for a circular junction of diameter $w$ = 1.0 $\mu$m we expect the first zero in the Airy pattern to occur at $\mu_0 H \approx$ 12 mT, where we have used $d_N+d_F$ = 50 nm to account for all the ferromagnetic layers and Cu spacers in the junctions. So as long as the Airy shifts are much less than 12 mT, we can ignore them.

A final consideration in the sample design is the number of magnetic layers in the junctions that are patterned during the ion milling fabrication step, as opposed to being left as extended thin films.  We have fabricated samples with various combinations of ion mill depths.  In general, we found that patterned SAFs were softer than unpatterned SAFs; hence in the samples reported here, only the top NiFe layer is patterned while both the SAF and the bottom Ni layer remain as extended films.  The final Josephson junction samples had the following layer structure: Nb(100)/Cu(5)/Ni(1.2)/Cu(10)/Co(4)/Ru(0.75)/Co(4)/
Cu(10)/NiFe(1.0)/Cu(5)/Nb(20)/Au(15)/Nb(150)/Au(20), where all thicknesses are in nm.  Details of the sample fabrication procedure are provided in \cite{Niedzielski:15}.

\begin{figure}[tbh!]
\begin{center}
\includegraphics[width=3.2in]{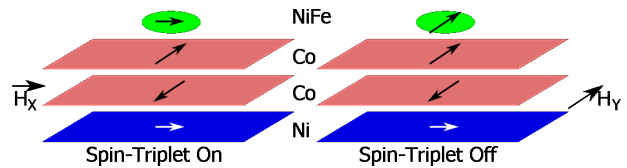}
\end{center}
\caption{(color online) Schematic diagram of the magnetization directions for the ferromagnetic layers in our Josephson junctions in the spin-triplet supercurrent ``on" state (left) and ``off" state (right).  The sample is initialized into the on state by a large field, $\mu_0 H$ = 260 mT, in the longitudinal (X) direction.  Thereafter, only small applied fields, $\mu_0 H <$ 20 mT, are applied in either the X or Y directions to rotate the magnetization of the soft NiFe layer while leaving the magnetizations of the hard Ni layer and Co/Ru/Co SAF unchanged. }\label{SchematicDiagram}
\end{figure}

Completed samples were mounted on a cryo-probe and measured at 4.2 K in a liquid helium dewar equipped with a Cryoperm magnetic shield. The probe had a pair of orthogonal coils to provide magnetic fields at any direction in the sample plane.  To initialize the magnetization of each layer, a large (260 mT) magnetic field is applied longitudinally along the sample, i.e. along the x direction shown in the left side of Figure 1. This causes the Ni and NiFe magnetizations to align in the x-direction while the two Co layers in the SAF point along the $\pm$y-directions after the spin-flop transition and subsequent reduction of field, creating orthogonality of the magnetizations to optimize generation of spin-triplet supercurrent.  To remove trapped flux in the Nb electrodes, the sample was briefly lifted to just above the liquid He level in the dewar, then re-immersed.  Current-voltage characteristic curves were measured using a 4-terminal SQUID-based self-balancing potentiometer circuit.  The current was stepped from zero well past the critical current in each direction; the curves were fit by the standard form for overdamped Josephson junctions: $V=R_N(I^2-I_c^2)^{1/2}$, where $R_N$ is the normal state resistance and $I_c$ is the critical current.  By repeating this measurement while iteratively stepping $H_x$ through a field range typically -20 to 20 mT, an Airy pattern is measured in the longitudinal direction.  This initial Airy measurement serves as a test of junction quality.

\begin{figure}[tbh!]
\begin{center}
\includegraphics[width=2.6in]{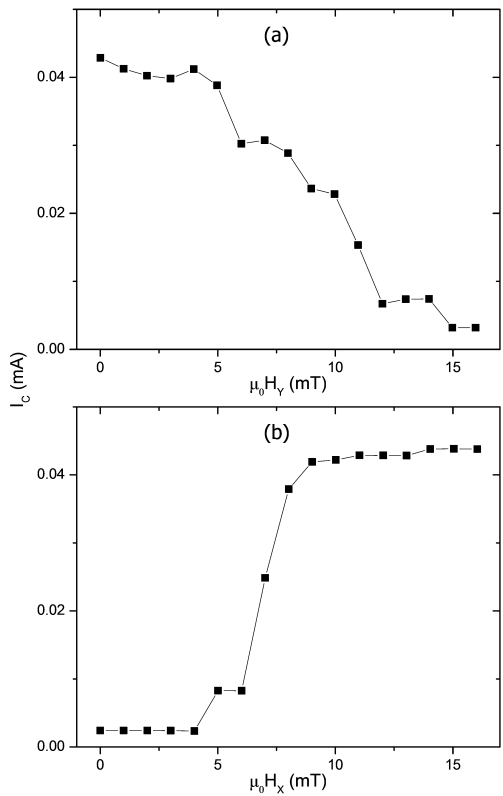}
\end{center}
\caption{Evolution of the Josephson critical current, $I_c$, of sample $\#$2, as a function of the magnitude of the ``set" field.  (a) Field applied in the transverse (Y) direction turns spin-triplet supercurrent off. (b) Field applied in the longitudinal (X) direction turns triplet supercurrent back on. All measurements are made in zero field.}\label{TurnOnTurnOff}
\end{figure}

From here, a small transverse magnetic field $H_y$ is applied and removed, and $I_c$ is measured again at $H$ = 0.  This process is repeated with increasing values of $H_y$.  The results are shown in Figure 2(a) for sample $\#$2.  As seen in the figure, $I_c$ starts at about 45 $\mu$A, then decreases with increasing $H_y$ until it saturates at the low value of $\approx$ 3 $\mu$A for $\mu_0 H_y$ = 16 mT.  This decrease is due to rotation of the NiFe magnetization until it aligns nearly collinearly with the Co layers in the SAF, thereby turning off the spin-triplet-generation mechanism.  The process can be reversed by applying gradually increasing values of $H_x$, as shown in Figure 2(b) -- again with all measurements performed at $H$ = 0.  For this sample, $I_c$ returns to its initial value when $\mu_0 H_x$ reaches about 10 mT.  The fields required to turn the triplet supercurrent on and off vary somewhat from sample to sample; hence the initial procedure described in Figure 2 determines the magnitude of the maximum external field that will be used for all ensuing measurements.  If we apply too large a field during the ``turn-off" step shown in Figure 2(a), then the supercurrent starts to increase again, indicating that the SAF is starting to rotate.  If that happens, the sample must be re-initialized and the experiment started over again.

\begin{figure}[tbh!]
\begin{center}
\includegraphics[width=2.8in]{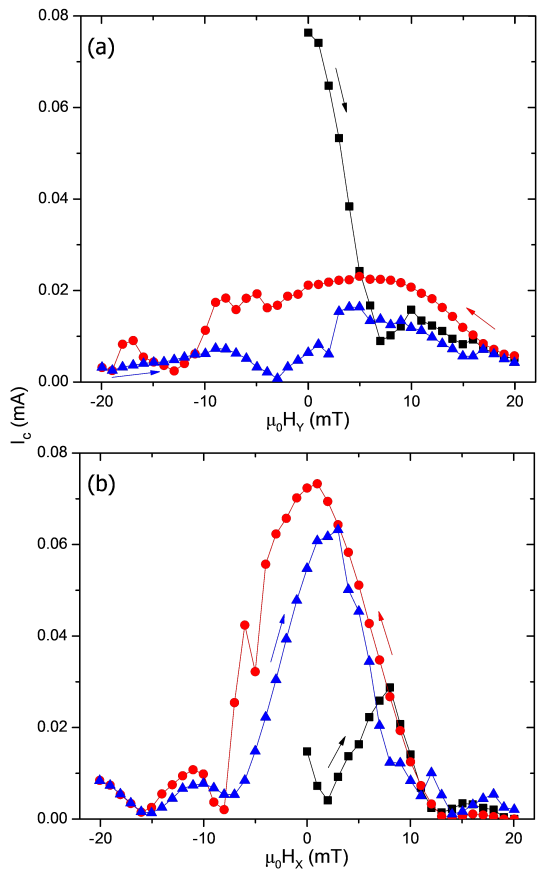}
\end{center}
\caption{(color online)  Plots of critical current vs applied field, known as Fraunhofer patterns.  (a) The sample starts in the ``on" state at H = 0, with $I_c$ = 76 $\mu$A.  As a field is applied in the transverse (Y) direction (black squares), $I_c$ drops rapidly due to both the Fraunhofer physics and the turning off of the spin-triplet supercurrent.  To verify the latter, the transverse field is swept from +20 to -20 mT (red circles) and then back to +20 mT (blue triangles).  The supercurrent stays low throughout this entire transverse Fraunhofer pattern.  (b) The sample starts in the off state at H = 0, with $I_c \approx 15 \mu$A.  As a longitudinal (X) field is applied, $I_c$ starts to increase due to the spin-triplet supercurrent turning on, but immediately drops due to Fraunhofer physics.  To verify the former, the longitudinal field is swept from +20 to -20 mT (red circles) and then back to +20 mT (blue triangles).  The Fraunhofer pattern exhibits a maximum supercurrent close to the initial value of 76 $\mu$A.}\label{Fraunhofers}
\end{figure}

To further test the robustness of these observations, we carried out measurements of $I_c$ in the presence of the applied field, i.e. measurements of the longitudinal and transverse Airy patterns in the on and off states, respectively.  Figure 3(a) shows the results for sample $\#$1, with  $\mu_0 H_{max}$ = 20 mT.  In the initial on state, sample 1 has $I_c \approx 76 \mu$A.  We first ramp the transverse field $H_y$ from 0 to $H_{max}$, and $I_c$ drops precipitously with increasing $H_y$, as expected.  Not only does the field rotate the magnetization of the NiFe and therefore turn the spin-triplet supercurrent off, but the presence of the field also causes $I_c$ to decrease as the first minimum in the Airy pattern is approached.  To separate the two effects, a full Airy pattern is now measured as a function of field $H_y$ in the transverse direction: $H_{max} \rightarrow -H_{max} \rightarrow$ $H_{max}$.  Figure 3(a) shows that $I_c$ remains low during this process, confirming that the spin-triplet supercurrent remains suppressed for all fields. The whole process is now repeated while applying a longitudinal field $H_x$.  After an initial transient, Figure 3(b) shows that $I_c$ starts to rise for 2 mT $< \mu_0 H_x <$ 8 mT, but then drops as $H_x$ approaches the location of the first minimum in the Airy pattern.   During the subsequent full longitudinal field sweeps, $H_{max} \rightarrow -H_{max} \rightarrow$ $H_{max}$, $I_c$ exhibits a large-amplitude Airy pattern with maximum value close to the initial value of $I_c$ = 76 $\mu$A, demonstrating that the spin-triplet supercurrent has been turned back on. Figure 3(b) shows that there is some shift in the central peak position of the Airy pattern due to the changing NiFe magnetization direction, but the effect is not large enough to obscure the results.

\begin{figure}[tbh!]
\begin{center}
\includegraphics[width=3.2in]{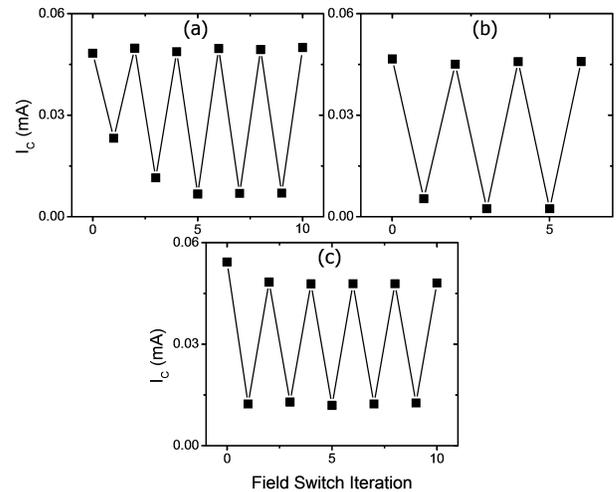}
\end{center}
\caption{ Demonstration of ``on-off" switching of the spin-triplet supercurrent.  Each figure shows a different sample as fields of $H_{max}$ are applied alternately in the transverse and longitudinal directions, for odd and even values of the field iteration, respectively.  The values of $H_{max}$ used for these three samples were 20 mT, 16 mT, and 10 mT for (a), (b), and (c), respectively.}\label{OnOffResults}
\end{figure}

Finally, we test the repeatability of the result by applying alternately fields in the two directions: $H_y$ = $H_{max}$ and $H_x$ = $H_{max}$, measuring $I_c$ in zero field for each iteration.  The results are plotted in Figure 4 for all three samples.  After an initial transient, which is not understood, all samples show repeatable behavior, with $I_c^{on}$/$I_c^{off}$ $\approx$ 7, 19, and 5 for samples 1 - 3, respectively.  Note that the value of $I_c$ in the on state varies little between the three samples, whereas the off-state values vary considerably due to small uncontrolled misalignments of the layer magnetizations that produce residual spin-triplet supercurrent.

In conclusion, we have demonstrated control of the amplitude of spin-triplet supercurrent through manipulation of the magnetization direction of neighboring F layers in multi-ferromagnet S/F/S Josephson junctions. In addition, we have shown that the states are stable and maintained when the magnetic field is turned off, and reproducible over multiple iterations and in multiple samples.  This work represents a major step forward in the control of spin-triplet correlations induced in superconducting/ferromagnetic heterostructures, hence in the development of the nascent field of superconducting spintronics \cite{Eschrig:11,Linder:15}.

Acknowledgements: We thank B. Bi, A. Cramer, and R. Loloee for technical support, and B. Niedzielski for a critical reading of the manuscript.  The work reported here was funded by the U.S. Department of Energy, Office of Basic Energy Sciences, Division of Materials Sciences and Engineering under Award DEFG02-06ER46341.  Sample fabrication was carried out in the Keck Microfabrication Facility at Michigan State University.

\end{document}